# Ultra-slow capillary rise on hydrogel surfaces

Anagha Datar,[a] Joonas Ryssy,[a] Aku O. Toivonen,[a] and Matilda Backholm*[a]

Capillary rise occurs when a thin tube contacts a liquid, which rises against gravity due to the capillary force. This phenomenon is present in a wide range of everyday and industrial settings and provides the means to measure the physical properties of liquids. Here, we report on the unusual ultra-slow capillary rise on a solid-like material of agarose hydrogels. The observed meniscus motion cannot be described with classical capillary rise models, and we develop a new model based on the fluid transport through the porous hydrogel network. Our model is in good agreement with the temporal scaling observed in our experiments with agarose gels made with five different concentrations and with two different viscosities of the liquid flowing inside the gel. Our results provide a non-invasive technique to directly estimate the permeability of hydrogel interfaces with high spatial resolution, which is important in the implementation of hydrogels in advanced biomedical applications.

## 1. Introduction

Hydrogels consist of crosslinked polymer networks that can absorb and retain large amounts of water or biological fluids, resembling the structure of living tissues[1]. These soft materials have become essential in a wide range of biomedical applications[2,3] showcasing valuable self-healing properties[4]. The permeability of hydrogels, that is, the bulk fluid flow through them, affects their mechanical properties[5], while the interfacial fluid transport affects their adhesivity[6–8]. The latter is crucial for the friction of hydrogel surfaces[9] and the implementation of hydrogel interfaces in bioadhesion applications[3,10–12], where product development is complicated by the wet conditions of the interface[13]. Hydrogel interfaces are also commonly used in the plant root growth research community to mimic soil[14–18], and the interfacial and bulk fluid transport capacity of these model systems is of great agricultural and ecological importance[16]. Finally, the pore size and connectivity (i.e., the permeability) of hydrogels directly influences the 3D migration of cells through the gel[1], whereas the gradients in the substrate stiffness influence cell mobility on the hydrogel surface[19]. It remains unclear if and how the varying interfacial permeability on such a substrate plays a role in the 2D cell migration. The bottleneck in this development lies in probing the interfacial fluid transport properties with spatial resolution down to the length scale of single cells.

The mechanism of fluid transport through the bulk of the porous polymer network of a hydrogel can be described using Darcy's law[20,21] $Q = (\kappa A/\eta L)\Delta p$, where $Q$ is the volumetric flow rate, $\kappa$ the Darcy permeability (units of $\mathrm{m}^2$), $A$ the cross-sectional area of the material through which the fluid is flowing, $\eta$ the dynamic viscosity of the fluid, and $L$ the distance over which the pressure drop $\Delta p$ is calculated. Bulk permeability, that is, the ability of fluids to pass through a large chunk of material, can change by orders of magnitude for different porous materials[22] and is important for hydrogel-based biomedical applications[2] such as drug delivery, tissue engineering, lubricants, and contact lenses. Measuring $\kappa$ is often challenging due to its intrinsic nature as well as the ease of gel compression during the test[23], but has been achieved using, for example, diffusion experiments[24], pressure-driven flow methods[25], and compression experiments[26]. An empirical equation relates the volume fraction of water ($\varphi^w$) with the Darcy permeability of agarose and cartilage as[26] $\kappa = 0.0039(\varphi^w/1-\varphi^w)^{3.236}$ (nm$^2$). However, measuring $\varphi^w$ requires a time-consuming gravimetric approach where the wet mass of a hydrogel piece is compared with the dry mass after 20 h of oven drying[27]. Alternatively, the permeability can be empirically determined as $\kappa = C\left(2r_{\mathrm{p}}\right)^2$ based on the pore size $r_{\mathrm{p}}$ of the hydrogel, where $C$ is a dimensionless constant[23]. However, this requires advanced and expensive CRYO-SEM measurements[28], or cumbersome techniques such as mercury intrusion[29] and capillary flow porometry[30], to name a few. Most of these experimental approaches are invasive, with effects of freezing, flow of other liquids than water, or confinement into cylindrical chambers that can strongly alter the size and/or distribution of the pores. Furthermore, for capillary force-driven adhesion applications and 2D cell migration studies of hydrogels, the spatially resolved measurement of fluid transport is of greater relevance. However, none of the conventional techniques can probe permeability with high spatial resolution in a non-invasive manner.

Capillary rise is a classical tool in fluid mechanics to probe liquid properties[31–33] (**Fig. 1a, Supplementary Movie S1**). Here, a hollow cylinder (inner radius $R$) is placed in contact with a liquid (surface tension $\gamma$, density $\rho$, viscosity $\eta$) which rises in the cylinder due to the capillary force $F_\gamma \sim R\gamma\cos\theta$, where $\theta$ is the (dynamic[34–36]) contact angle between the liquid and the inner cylinder wall. For viscous fluids, damping is dominated by viscous forces[37] $F_\eta \sim \eta z \frac{\mathrm{d}z}{\mathrm{d}t}$, where $z$ is the meniscus position. Balancing the capillary and viscous force renders the Lucas-Washburn equation[32,38] $z(t) \sim (\gamma\cos\theta\, R/\eta)^{1/2}\, t^{1/2}$. For lower viscosity fluids, the initial damping is dominated by inertial forces $F_{\mathrm{i}} \sim \rho R^2 \left(\frac{\mathrm{d}z}{\mathrm{d}t}\right)^2$ and the capillary rise dynamics becomes linear[37,39,40] $z(t) \sim (\gamma\cos\theta/\rho R)^{1/2}\, t$. The rise of the meniscus equilibrates at a constant height as the weight of the liquid column balances the capillary force, as given by Jurin's law[31,33] $h_{\mathrm{eq}} = 2\gamma\cos\theta/\rho g R$. By tracking $z(t)$ and $h_{\mathrm{eq}}$ of the capillary rise, the bulk and/or interfacial properties of the fluid can be probed[41]. Studies on capillary rise in soft capillaries[42] and flexible structures[43] in contact with liquid has shown interesting deviations from these classical models.

Here, we have performed capillary rise experiments with glass microcapillaries in contact with the interface of a solid-like material represented by agarose hydrogels (**Fig. 1b, Supplementary Movie S2**). An unexpected, ultra-slow capillary rise of fluid from the gel was observed that cannot be explained with classical capillary rise models. We developed a new model based on the fluid flow transport through the hydrogel pores to capture the experimental observations. Our experiments render quantitative measurements of the flow across agarose interfaces with spatial resolution and represent a simple, fast, and non-invasive method to directly measure hydrogel permeability. This is important for the current surge in research for developing hydrogel interfaces perfected for different biomedical applications.

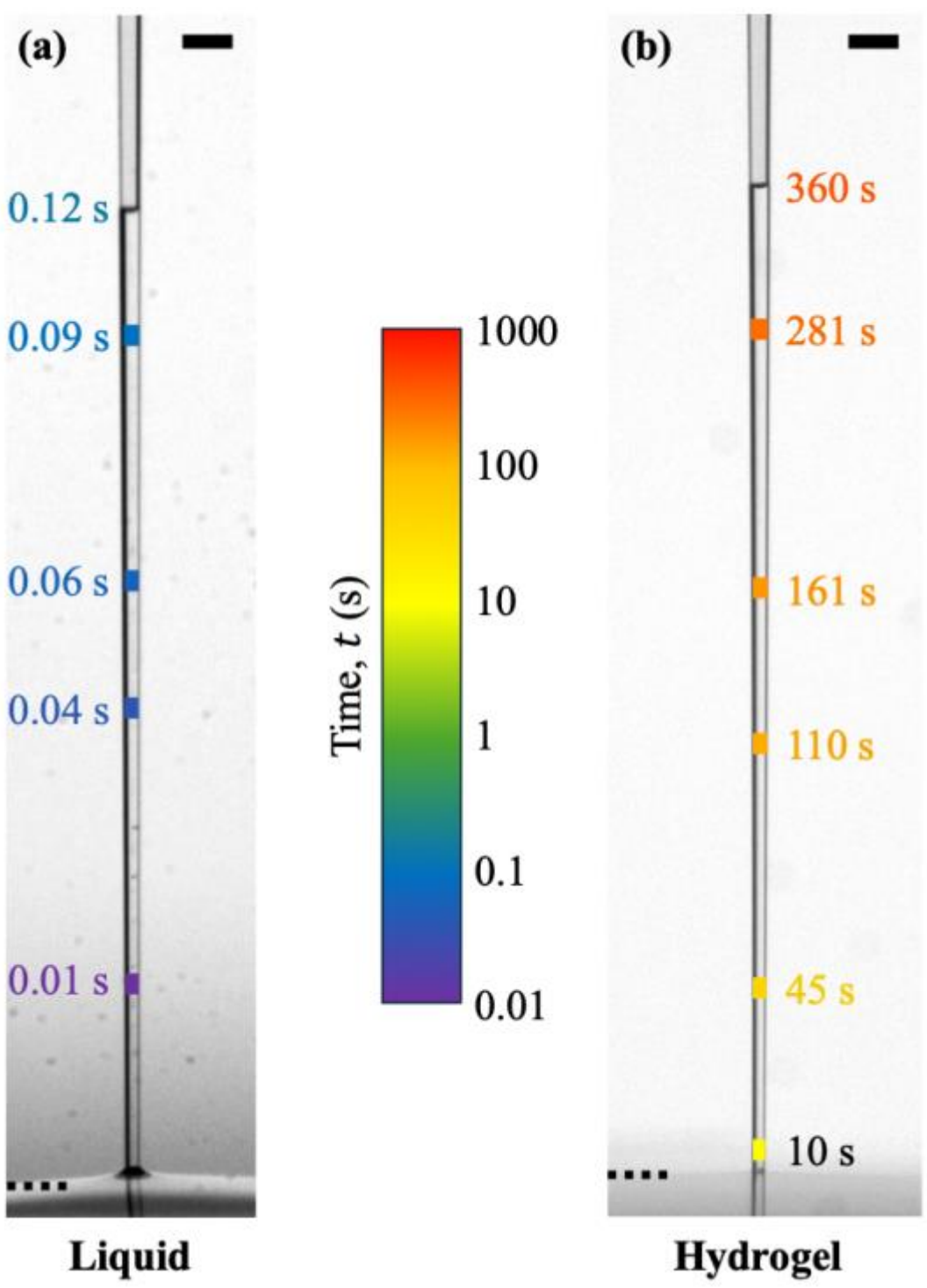

**Fig. 1 | Microcapillary rise on a liquid and a solid-like material. a)** Image sequence of the rapid meniscus rise in a glass microcapillary (inner tip radius $R = 22.9 \pm 0.6$ µm) when contacting a puddle of the water-like liquid that has leaked out of an agarose hydrogel. **b)** Image sequence of the ultra-slow meniscus rise in a glass microcapillary ($R = 20 \pm 3$ µm) in contact with an agarose hydrogel. The time is measured from the point of contact with the material for both cases. Due to the non-linear tapering of the microcapillary, analysis of the rise velocity is done only in the early regime (rise height $z < 500$ µm) where the microcapillary radius is constant. The dashed black line indicates the interface. Scale bar 200 µm. The technical details of this 0.5% agarose$^{++}$ sample are described in detail below.

## 2. Materials and Methods

### 2.1 Microcapillary manufacturing

We followed the micropipette force sensor protocol[44] when manufacturing the microcapillaries. The microcapillaries were pulled out of 1 mm thick glass capillaries with an inner diameter of 0.75 mm (World Precision Instruments, model no. TW100-6) using a micropipette puller (Narishige, model no. PN-31). The pulled end of the microcapillary was cut with a microforge (Narishige, model no. MF-900) to length of 11-14 mm. The capillary radius was measured from a cross-sectional image taken with 1x (Mitutoyo 378-800-12 for unpulled capillaries) or 2x (Mitutoyo 378-801-12 for pulled microcapillaries) magnification objective. In the article, we have used the term 'capillary' for unpulled, uncut glass capillaries and the term 'microcapillary' for pulled and cut glass-capillaries.

### 2.2 Synthesis of hydrogels

Agarose hydrogels were prepared in two different ways. The first was prepared with only agarose and MES (2-Morpholinoethanesulfonic acid) buffer to balance the pH (gels referred to as agarose). The second was made with agarose, MES buffer, MS media (Murashige and Skoog Basal Medium) and sucrose (gels referred to as agarose$^{++}$). The latter is used extensively in the plant research community as artificial soil[14–18], and the fluid transport of these gels is of great importance[16]. The agarose (Invitrogen UltraPure™ LMP Agarose 16529) hydrogels were synthesized using a one-pot method. In both cases (agarose and agarose$^{++}$), a suitable w/v % of agarose powder was measured into a flask. In a similar fashion other chemicals were measured into the flask. For the agarose samples with plant growth media, these chemicals include MES monohydrate (Sigma-Aldrich 69892), Sucrose (Sigma-Aldrich S0389), and MS basal salt mixture (Sigma-Aldrich M5524). A required amount of DI-$H_2O$ (Sartorius Arium Mini) was added into the flask. The final concentrations were 1 w/v % Sucrose, 2.14 mg/mL MS salt, and 2.7 mM MES. The pH of the solution was adjusted to 5.7 with a small volume of 2M NaOH solution (Sigma-Aldrich 221465). The pH was confirmed using a pH meter (Mettler Toledo SevenDirect SD20). For the pure agarose sample, the previous protocol was followed sans MS salt mixture and sucrose. Both solutions were heated until boiling in a microwave (Whirlpool MAX). Upon boiling, the solution was quickly casted onto a petri dish and was allowed set in room temperature. The gel was stored at 4°C (Liebherr Mediline).

Hydrogel samples with a concentration gradient were prepared in the following manner. Two agarose solutions were made, 2% w/v and 0.5% w/v respectively. The gels were cast into polystyrene containers (74 x 43 x 5 mm), which were chilled on top of flat ice packs. The agarose was melted into the solution with microwave assisted heating. While both solutions remained hot, they were poured simultaneously into each end of the container, slightly mixing in the middle of the cooled containers. The containers were kept on top of the ice packs until the gels were set. The solidified gels were finally placed in large rectangular petri dishes with closed lids and stored at 4°C.

### 2.3 Capillary rise experiments on hydrogel interfaces

The experimental setup was built on an active vibration isolation stage (MVIS 30×30 model by Newport Corporation). A rectangular piece of size ca. 1 cm x 2 cm was cut out from the hydrogel (set in a 10 cm diameter plastic petri dish), placed flat on a clean glass coverslip and mounted on a flat aluminum plate acting as the imaging stage connected to a linear motor (Thorlabs MT1/M-Z9 - 12 mm). A microcapillary (aperture radius in the range of $R = 16 - 32$ µm) was held vertically above the gel, using a pole (Dynamically Damped Post, 14" Long, Metric, O1.5", Thorlabs, DP14A/M), an *xyz*-manipulator (MN-153, Narishige) and a pipette holder (IM-H3 Injection Holder Set by Narishige). A new, dry microcapillary was used for each hydrogel experiment to avoid unwanted effects on the flow from dried agarose fluid in the capillary. In the beginning of a recording, the microcapillary tip was placed at a short distance of about 50 µm from the upper surface of the gel using the micromanipulator. The linear motor was then triggered to move the sample towards the microcapillary at the speed of 5 µm/s. The motor was manually stopped when the contact was noticed, for all gels except 2% agarose$^{++}$. For this stiffest gel, the micropipette was allowed to push in a bit further, for maximum 50 µm to let the fluid flow happen. An extra micropipette, fixed to the imaging stage was used as a reference to measure the extent of stage movement and to keep a check on the amount of indentation of the microcapillary onto the hydrogel. Imaging was done at 60 fps (except on the 2% gels where 1 fps was enough) with a 2x Mitutoyo objective connected to a horizontal video microscopy unit (Mitutoyo 378-506, VIS-NIR range) and a small camera (Integrated Imaging Solutions, Inc. FLIR GS3-U3-23S6M-C). Recording was started a few seconds before the microcapillary-gel contact happened and ended typically when the meniscus went out of the field of view.

### 2.4 Sampling fluid from hydrogel

Small volumes of fluid leaked out from the three softer agarose gels (0.3, 0.5, and 0.7%). No such fluid could be collected from the stiffer gels. A few wells of 1 cm x 1 cm hole were cut out from the gel set in a 10 cm Petri dish, resulting in formation of puddles of fluid leaking out from the gel. The rate of fluid outflow went from high to low as the concentration of the hydrogels increased. After cutting out the wells, the Petri dish lid was closed. After waiting for a few hours, enough liquid was collected to form 200 µL drops for the capillary rise trials and rheology. The average density of the liquid samples was measured by weighing 50 µL volumes (measured with calibrated micropipette, Thermo Scientific™ Finnpipette™ F2 4642070) of the samples with a milligram balance (Fisher Scientific PAS214C).

### 2.5 Capillary rise experiments on fluid from hydrogel

A drop of at least 200 µL was placed on a clean glass coverslip which was placed on the imaging stage just like in the experiments on hydrogels. A glass capillary ($R = 0.335 \pm 0.011$ mm) was held directly above the drop as described earlier and the experiment was carried out in the same way as for hydrogels except that imaging was done at 500–800 fps using a 1x objective along with a fast camera (AOS LPRI1000, AOS Technologies AG, Switzerland). Recording was started a few seconds before the capillary-liquid contact happened and typically ended a minute after the meniscus had stabilized at its final height.

### 2.6 Rheology

Cone-plate rheology (Anton Paar MCR301, probe CP25-2/TG) measurements were performed on the above described oozed-out fluid from the 0.5% agarose and 3 softest agarose$^{++}$ (0.3, 0.5, and 0.7%) gels. Within the shear rate range accessible with the rheometer for this low water-like viscosity fluid (152–1000 $s^{-1}$), the viscosity of all agarose$^{++}$ samples remained constant at the viscosity of $\eta = 1.96 \pm 0.22$ mPa s (**Supplementary Fig. S1**). Three measurements were performed per concentration, and within error, there was no difference between the three different samples. Similarly, the viscosity of the fluid from three 0.5% agarose samples were measured as $\eta = 1.12 \pm 0.01$ mPa s. The fluid from all samples was assumed to consist of a low concentration of mostly short and thereby unentangled polymers, since most long polymers would be bound to the hydrogel network. The fluid was thus assumed to remain Newtonian also at the ultra-low average shear rates of the capillary rise experiments $U/R \approx 10^{-1}$ to $10^{1}$ $s^{-1}$.

### 2.7 Water content measurements

Wet masses of the hydrogels were measured on a balance (Fisher Scientific Precision Series) after the gels were set. After this, the gels were left to dry in a lateral flow hood (Thermo Scientific HERAguard HPH 12) under ambient conditions for a minimum of 72 hours. After the gels were fully dried, the dry masses were measured using the same balance. The water content was then calculated using[27] $\varphi^{w} = \frac{(m_{\mathrm{wet}} - m_{\mathrm{dry}})}{m_{\mathrm{wet}}} \cdot 100\%$ w/w.

### 2.8 Data analysis

The meniscus was tracked using a cross-correlation-based MATLAB code (deflection.m shared in Ref.[44]). The meniscus rise velocity ($U$) was calculated by making a linear fit to the $z(t)$ data when the meniscus had detached from the gel. Errors were estimated using the variance method.

## 3. Results and Discussion

**Microcapillary rise on a solid-like material**

Microcapillary rise experiments were performed on the agarose$^{++}$ gels with five different concentrations (0.3, 0.5, 0.7, 1, and 2 w/v%) as well as the 0.5 w/v% agarose gel. When contacting a hydrogel surface, ultra-slow capillary rise is observed (an example with 0.5% agarose$^{++}$ hydrogel is shown in **Fig. 1b**). The position of the meniscus was tracked as a function of time and was observed to follow a linear scaling for all samples (**Fig. 2a**). In these experiments, especially on the stiffer gels, some time was required for the meniscus to develop in the tip of the microcapillary and detach from the hydrogel. In **Fig. 2a**, $t$ and $z$ were zeroed when the meniscus had detached and started rising. The velocity of the meniscus decreases strongly with increasing agarose concertation (**Fig. 2b**).

A linear $z \sim t$ scaling is predicted in the inertial regime for normal capillary rise experiment: $z(t) \sim (\gamma \cos\theta / \rho R)^{1/2}\, t$. However, if this classical equation is fit to the experimental data, a highly unphysical value for the surface tension emerges, which is of the order of $\gamma \sim 10^{-12}$ $Nm^{-1}$. For comparison, the surface tension of water is $\gamma \sim 10^{-1}$ $Nm^{-1}$, while the best surfactants[45] achieve $\gamma \sim 10^{-4}$ $Nm^{-1}$ and some phase separated systems[46] go as low as $\gamma \sim 10^{-6}$ $Nm^{-1}$. The liquid that flows through a hydrogel is primarily composed of water, so it cannot have a surface tension eleven orders of magnitude lower than that of water. It is well known that surface tension measurements using the capillary rise technique can be affected by factors such as whether the capillary is dry or pre-wet[37], the cross-sectional geometry of the capillary[47,48], the potentially complex fluid properties of the liquid[49], as well as the dynamic contact angle during the rise[35]. However, our ultra-slow hydrogel capillary rise results cannot be explained by the implementation of a dynamic contact angle since the rise is so slow (Capillary number $\mathrm{Ca} = \eta U/\gamma \sim 10^{-8}$ to $10^{-6}$) that the system can be considered quasi-static[35]. Furthermore, the results are not altered much using a pre-wet microcapillary (**Supplementary Fig. S2**) nor would the slightly tapered geometry of the microcapillaries cause a decrease in the velocity[48]. Our measurements of density and viscosity of the liquid leaking out of these gels show that the fluid is water-like and close to Newtonian. Finally, the Reynolds number of our system $\mathrm{Re} = RU\rho/\eta \approx 10^{-6}$ to $10^{-3} \ll 1$, which indicates that inertia is not the dominating force in the experiments. There is, in other words, another physical factor dampening the fluid flow in these hydrogel microcapillary rise experiments.

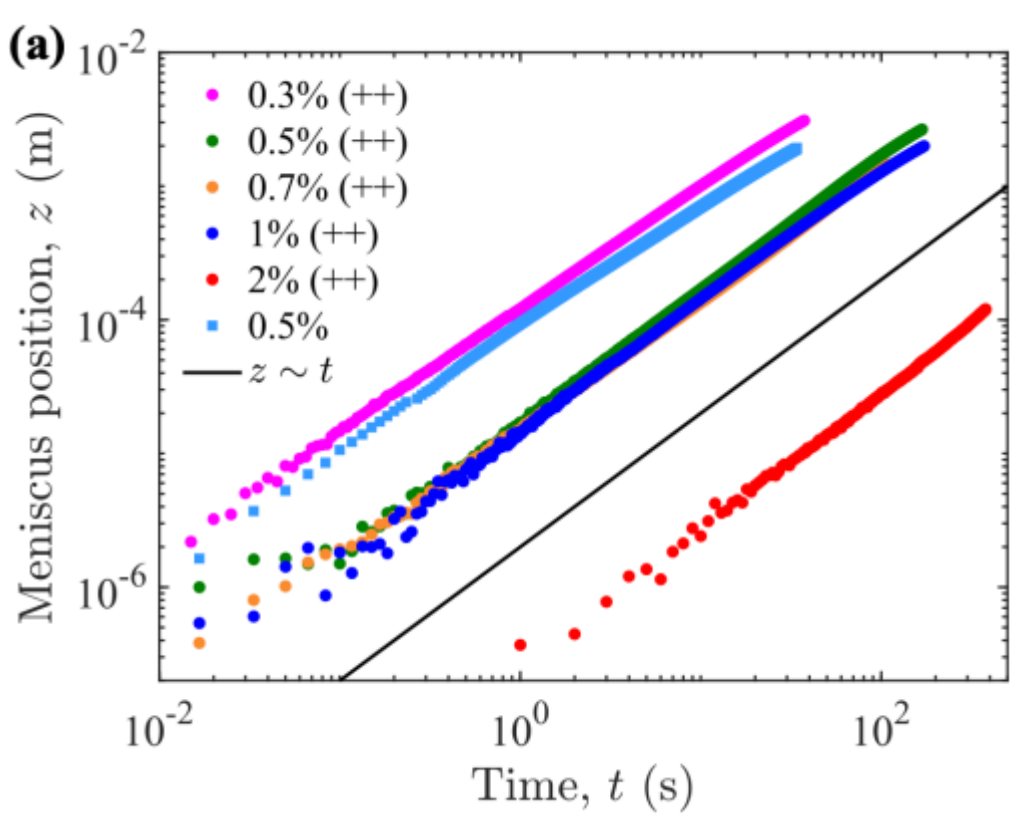

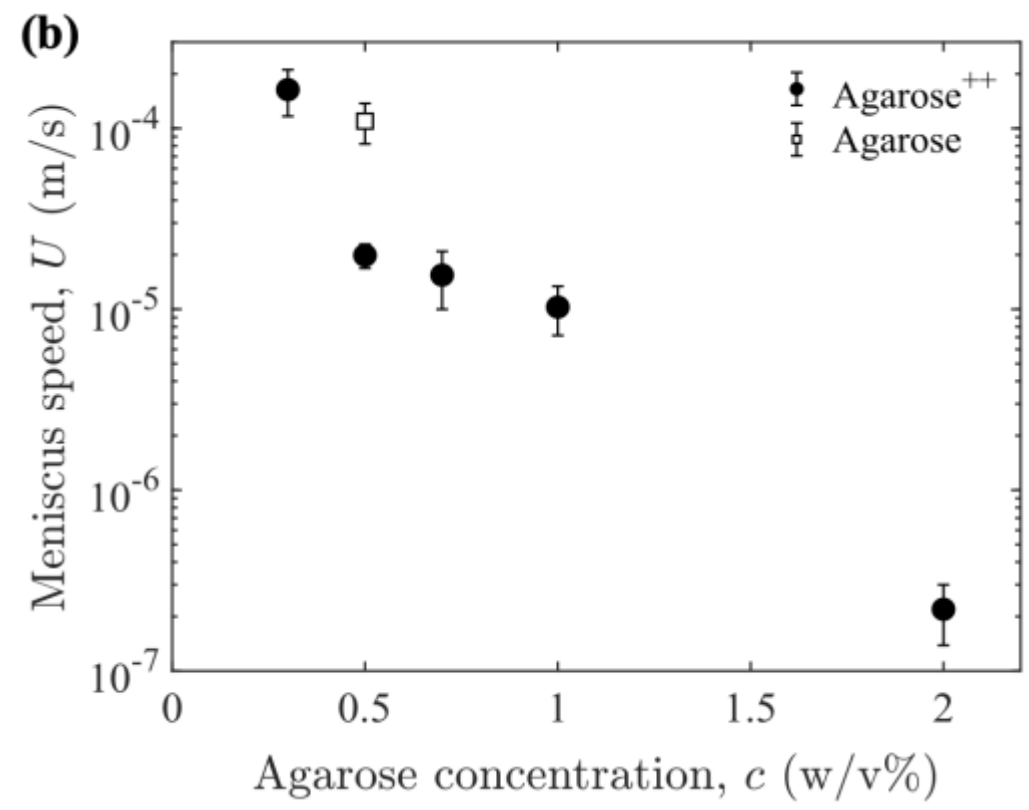

**Fig. 2 | Microcapillary rise on agarose hydrogels. a**) Meniscus position as a function of time in microcapillary rise experiments on the 0.5% agarose and different concentrations of agarose$^{++}$ samples with similarly sized glass microcapillaries ($R = 20.0 \pm 0.9$ μm). The solid line is a visual guide showing a slope of one, which indicates that the meniscus in all datasets rises at a constant speed in this loglog graph. The meniscus speed is determined by fitting a linear function $z = U \cdot t$ to the data. **b)** Average meniscus speed from many experiments as a function of agarose concentration. The error is the standard deviation.

**Fluid transport in hydrogels during capillary rise**

To describe our experimental observations, the damped flow rate through the porous hydrogel material was modelled using Darcy's law. The volumetric flow rate in the microcapillary is $Q = \frac{\mathrm{d}V}{\mathrm{d}t} = \pi R^2 \frac{\mathrm{d}z}{\mathrm{d}t}$, where $V$ is the volume of the fluid inside the capillary. Solving this for $z$ by using Darcy's law ($Q = \kappa A \Delta p / \eta L$) together with the capillary pressure as the driving pressure $\Delta p = 2\gamma \cos\theta / R$ gives

$$z(t) \sim \frac{2\kappa\gamma\cos\theta}{\eta R^2} t \equiv U_\mathrm{t} t, \qquad (1)$$

where $U_\mathrm{t}$ is the theoretical velocity of the meniscus. Here, the pressure has been assumed to drop over a distance of $L \sim R$ for a sample with thickness much larger than the micropipette radius. The flow is furthermore approximated to only occur through the slab of volume directly under the inner cross-sectional area ($A = \pi R^2$) of the capillary. Eq. (1) supports the observed linear scaling of the capillary rise on hydrogel surfaces (**Fig. 2a**). This scaling robustly holds over 2 orders of magnitude in time (see **Supplementary Fig. S3** for data of the later time regime). From Eq. (1), it can be noted that capillary rise on a typical 1% agarose surface is easily observable only with very thin capillaries, since a $R = 1$ mm capillary would render a rise velocity of ca. 10 nm/s, which would be very difficult to detect.

The same equation can be derived by considering the damping force of the system. Here, the dominating damping is assumed to be viscous and occur inside the cylindrical pores of the hydrogels, which can be written as $F_{\eta,\mathrm{p}} \sim A_\mathrm{p} \eta \frac{\mathrm{d}u}{\mathrm{d}r} \sim (R^3\eta / r_\mathrm{p}^2) \frac{\mathrm{d}z}{\mathrm{d}t}$, where $u$ is the velocity profile inside the pores, $A_\mathrm{p} \sim N_\mathrm{p} R r_\mathrm{p}$ is the (vertical) cross-sectional area of all pores over which the pressure is applied, $N_\mathrm{p} \sim R^2 / r_\mathrm{p}^2$ is the number of pores under the capillary, and $\frac{\mathrm{d}u}{\mathrm{d}r} \sim U / r_\mathrm{p} = \frac{1}{r_\mathrm{p}} \frac{\mathrm{d}z}{\mathrm{d}t}$. By balancing the capillary force $F_\gamma \sim R\gamma\cos\theta$ with $F_{\eta,\mathrm{p}}$ and solving the differential equation, we arrive at $z \sim \gamma\cos\theta\, r_\mathrm{p}^2 t / \eta R^2$, which is the same as Eq. (1) when using $\kappa \sim r_\mathrm{p}^2$.

It is important to note that the significant difference between this system and a classical capillary rise experiment with a normal liquid is that the dominating dissipation now mainly occurs in the hydrogel pores instead of inside the glass capillary. Here, we assume a rigid porous media although some dissipation will arise from the elastic deformation of the hydrogel network. Due to the very narrow pores, the hydrogel system renders a much higher dissipation and a very low capillary rise speed. The classical viscous dissipation inside the capillary will become dominating at meniscus heights of $z > R^3 / r_\mathrm{p}^2 \approx 0.8$ to 0.03 m for typical pore sizes of $r_\mathrm{p} = 100 - 500$ nm and $R = 20$ $\mu$m (calculated from when $F_\eta > F_{\eta,\mathrm{p}}$). This is beyond the micropipette lengths used in our work.

**Standard measurements of hydrogel properties**

To check the quantitative agreement between Eq. (1) and our experimental observations, we measured $\rho$, $\eta$, $\gamma$, $\theta$ and $\kappa$ using conventional techniques. The viscosity of the fluid from the $0.3 - 0.7\%$ agarose$^{++}$ and 0.5% agarose samples remained constant at $\eta = 1.96 \pm 0.22$ mPa s and $\eta = 1.12 \pm 0.03$ mPa s, respectively, for different shear rates. The viscosity of the fluid in the 1 and 2% agarose$^{++}$ samples was assumed to remain the same as in the softer agarose$^{++}$ samples. The average fluid density was measured as $\rho = 990 \pm 20$ kg/m$^3$.

To probe the surface tension and contact angle of the fluid flowing through the hydrogels, classical capillary rise experiments were performed on the puddles of liquid that had leaked out from the three softest samples (**Fig. 3, Supplementary Movie S3**). This was done with a glass capillary (i.e., not a microcapillary) which allowed for the measurement of the equilibrium height as well as the static contact angle at equilibrium, giving $\theta = 53.8 \pm 4.1$ deg for agarose$^{++}$ and $\theta = 58.2 \pm 3.8$ deg for agarose. Using Jurin's law, the average surface tension of the fluid in the $0.3 - 0.7\%$ agarose$^{++}$ samples was determined as $\gamma = 0.04 \pm 0.01$ N/m and $\gamma = 0.05 \pm 0.01$ N/m for the 0.5% agarose sample. The average agarose$^{++}$ values were assumed to hold also for the 1 and 2% samples that did not leak fluid. These values are a bit lower than literature values on agarose-water mixtures before the sol-gel transition ($\gamma \approx$ 0.06 N/m)[50]. The Darcy permeability of the agarose hydrogels was determined with the empirical equation[26] $\kappa_{\varphi^\mathrm{w}} = 0.00339(\varphi^\mathrm{w}/1 - \varphi^\mathrm{w})^{3.236}$ (nm$^2$) by first measuring the water content $\varphi^\mathrm{w}$ of the gels. The results from these measurements are listed in **Table 1**.

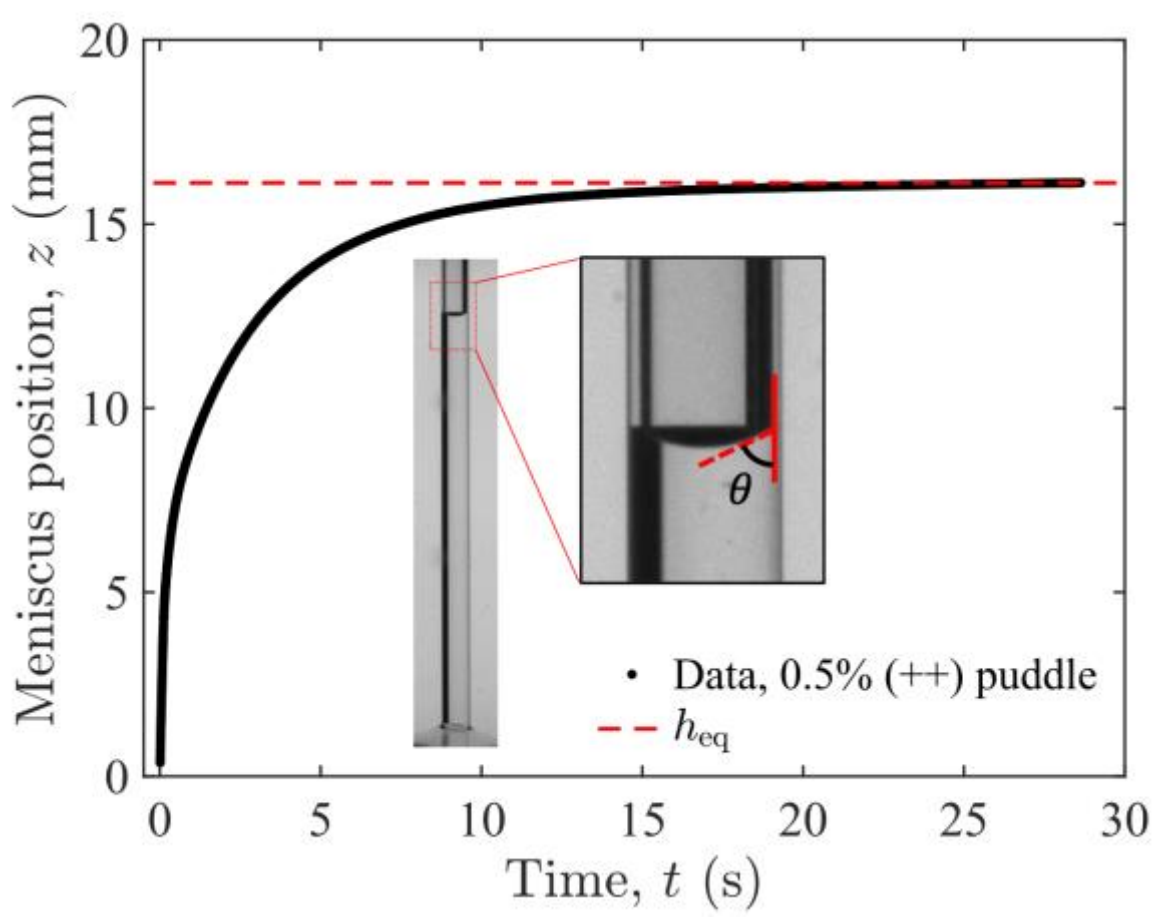

**Fig. 3 | Normal capillary rise on liquid puddles from agarose hydrogels.** Meniscus position as a function of time with the equilibrium height marked (dashed line) from an experiment on the liquid puddle that has leaked out from a 0.5% agarose$^{++}$ hydrogel. An image from the experiment at the final equilibrium position is shown in the inset with the static contact angle defined.

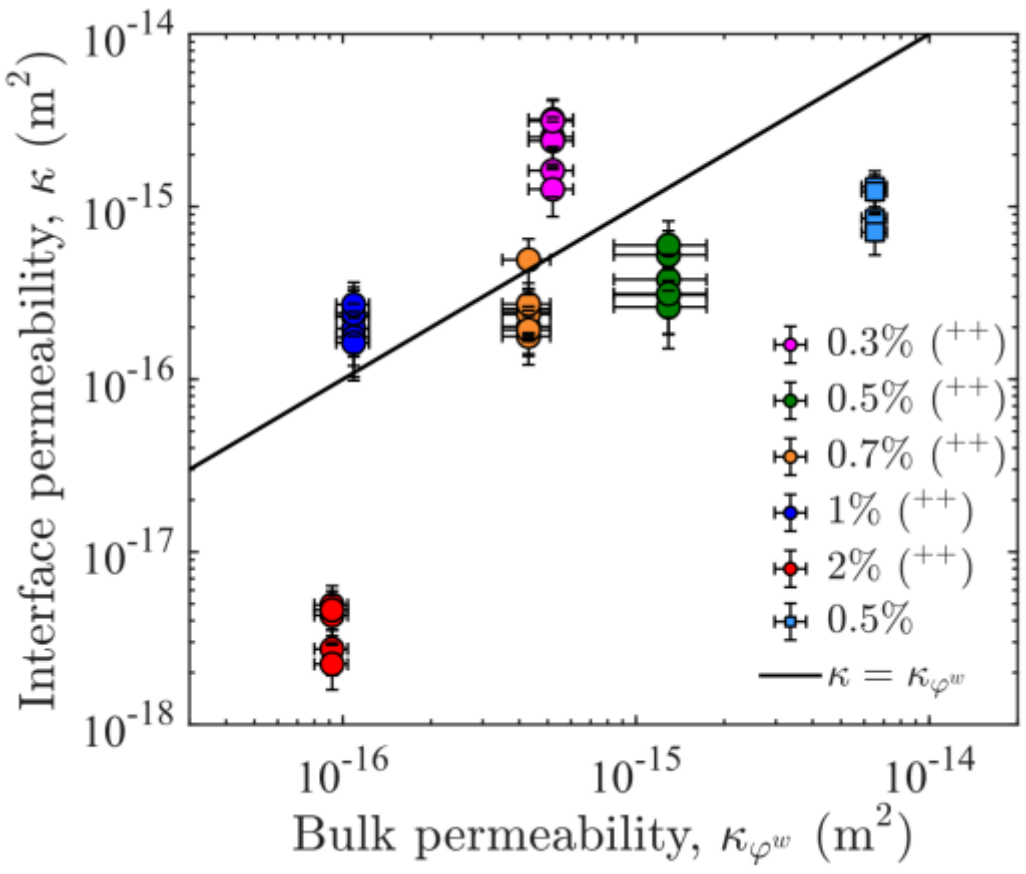

**Fig. 4 | Fluid transport across hydrogel interfaces.** Darcy permeability measured in the interfacial microcapillary rise ($\kappa$, calculated using Eq. (1)) versus bulk measurements based on the hydrogel water content ($\kappa_{\phi^w}$). The solid line has a slope of one and indicates where the interface and bulk permeabilities are the same. The errors are the error propagations of the components (and their standard deviations) in each equation for $\kappa$. Several experiments were performed on different locations of two independently made samples per concentration, and the different $\kappa$ values indicate the spatial variations of the sample. The solid line shows $\kappa = \kappa_{\varphi^w}$.

**Spatially resolved permeability measurements**

Using this experimental approach combined with Eq. (1), the interfacial permeability of agarose gels can be measured with the non-invasive method of microcapillary rise using $\kappa = U\eta R^2/2\gamma\cos\theta$. Here, the average values for $\eta$, $\gamma$, and $\theta$ have been included in the equation for each concentration together with $U$ and $R$ measured from experiments with different microcapillaries. Due to the ultra-slow meniscus motion in the hydrogel microcapillary rise experiments, these were assumed to be quasi-static, and we used the static contact angles (**Fig. 3**) as input for $\theta$ in Eq. (1). The bulk and interfacial permeabilities are known to be different[9], and directly probing the latter is important when designing hydrogel interfaces[9,13]. In **Fig. 4**, the measured $\kappa$ is against the bulk permeability $\kappa_{\varphi^w}$ determined through water content measurements. We find that the interface permeability aligns with the measured bulk values for the 0.3–0.7% samples. The scatter around the solid line in **Fig. 4** is attributed to inherent variations between different batches used for the microcapillary rise experiments and the water content measurements. In the stiffest 2% agarose$^{++}$ sample, the measured fluid flow across the interface is significantly lower than in the bulk. We assign this to an actual difference in the interface and bulk permeabilities of this stiff hydrogel. This highlights the importance of performing fluid flow transport measurements at the interface of the hydrogel if its purpose is to be used in, for example, a bioadhesion application. Hydrogel flow properties determined from measurements done in the bulk could be misleading for predicting the actual performance of the hydrogel interface.

**Table 1 |** Water content and bulk Darcy permeability of the agarose and agarose$^{++}$ hydrogels used in this work. The error for $\varphi^{\mathrm{W}}$ and $\kappa_{\varphi^w}$ is the standard deviation from measurements on 4–6 different plates and the propagated error, respectively.

| $c$ ($w/v$%) | $\varphi^w$ (%) | $\kappa_{\varphi^w}$ (m²) |
|---|---|---|
| 0.3 (++) | $97.56 \pm 0.13$ | $(5.2 \pm 0.9) \cdot 10^{-16}$ |
| 0.5 (++) | $98.2 \pm 0.2$ | $(12.9 \pm 4.5) \cdot 10^{-16}$ |
| 0.7 (++) | $97.42 \pm 0.15$ | $(4.3 \pm 0.8) \cdot 10^{-16}$ |
| 1 (++) | $96.11 \pm 0.15$ | $(1.09 \pm 0.14) \cdot 10^{-16}$ |
| 2 (++) | $95.9 \pm 0.2$ | $(0.92 \pm 0.12) \cdot 10^{-16}$ |
| 0.5 | $98.87 \pm 0.03$ | $(65 \pm 6) \cdot 10^{-16}$ |

Care should be taken to perform experiments that do not extract a noticeable fraction of liquid from the sample, as this would change the permeability over time. This is not a problem with thin microcapillaries that extract volumes of only ca. $0.1 \cdot 10^{-12}$ m³ (i.e., 0.1 nL) after a rise of $z = 100$ $\mu$m. The largest contribution to the error for $U_{\mathrm{t}}$ in Eq. (1) comes from the surface tension measurement (see **Supplementary Fig. S4**). Minimizing the error associated with measuring $\gamma$ would improve the sensitivity of this technique. Future work could focus on improved protocols to perform precise surface tension measurements with, for example, the pendant drop technique.

From our measurements, the effective pore size of the hydrogel can be deduced from the permeability as $r_{\mathrm{p}} \sim \kappa^{1/2}$. This can be a valuable avenue for estimating the hydrogel interface mesh size in a non-invasive manner. It should be noted that our model assumes a rigid network through which the fluid flows. An extended model considering the poroelasticity of the system would be an interesting future theoretical avenue to explore.

To highlight the potential of the microcapillary rise technique, we measured the permeability of an agarose sample made with a concentration gradient (**Fig. 5**). To the best of our knowledge, achieving spatial resolution is not possible with existing techniques. Here, the spatial resolution is at the length scale of the micropipette tip ($\sim 10^{-5}$ m), which is close to that of single cells. This shows the potential of microcapillary rise as a technique to probe intricate flow properties of structured hydrogel interfaces, which will be important in future bioadhesion and cell migration studies.

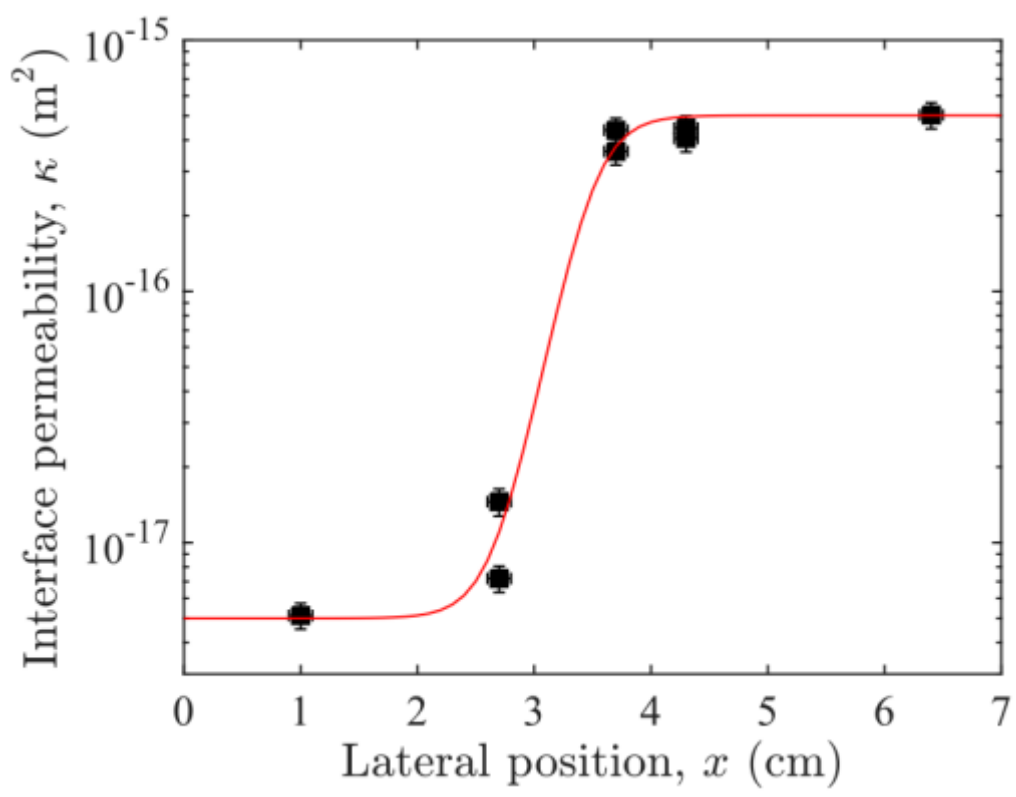

**Fig. 5 | Spatially resolved permeability of agarose with concentration gradient.** Permeability as a function of distance along the agarose substrate with a concentration gradient (2% to 0.5% from left to right). The red line is a sigmoid function ($\kappa = 5 \cdot 10^{-16}/(1 + e^{-5.5(x-3.5)}) + 5 \cdot 10^{-18}$) to guide the eye. The *y*-axis error is the error propagation of the components (and their standard deviations) in the equations for $\kappa$. The *x*-axis errors are the precision in determining the lateral position.

## Conclusions

Microcapillary rise experiments were performed on different agarose hydrogels. The constant, ultra-slow speed of meniscus rise was successfully modelled by implementing Darcy's law to describe the fluid transport through the hydrogel. This corresponds to dominating viscous dissipation in the small pores of the hydrogel and represents an interesting capillary phenomenon that can be described with a modified capillary rise model. The model holds for agarose gels with different pore sizes and different viscosities of the fluid flowing through.

We successfully show that the microcapillary rise technique can be used to directly measure the interface permeability of hydrogels in a non-invasive manner. This is a strong improvement compared to existing techniques, which often render gel compression[23], are based on an empirical water content formula[26] and long drying experiments[27], or are based on an empirical pore size equation[23] and cumbersome, expensive, and invasive measurements requiring sample freezing[28], flow of other liquids than water[29], or confinement of the hydrogel into cylindrical chambers[30], to name a few. All of these can strongly alter the size and structure of the porous hydrogel network and render incorrect measurements of the permeability. Furthermore, these probe the bulk permeability.

Interface permeability is a crucial intrinsic material property of hydrogels in the development of these for biomedical applications[2]. We show that the interface and bulk permeabilities can differ by an order of magnitude in a stiff agarose sample, highlighting the importance to directly probe the former if the material interface is to be used in an application. Furthermore, it is well documented that cell mobility is affected by gradients in the stiffness of hydrogel interfaces[19]. Our new microcapillary rise technique allows for interface permeability measurements with spatial resolution down to cellular length scales on such samples. This will be important in future work investigating how 2D cell migration is affected by changes in the hydrogel permeability.

## Author contributions

A.D.: investigation, methodology, resources, visualization, writing – review & editing; J.R.: investigation, methodology, resources, writing – review & editing; A.O.T.: investigation; M.B.: conceptualization, formal analysis, funding acquisition, investigation, methodology, project administration, supervision, validation, visualization, writing – original draft.

## Conflicts of interest

There are no conflicts to declare.

## Data availability

The datasets used for plotting all graphs in the paper as well as examples of raw data files are shared on Zenodo[51].

## Acknowledgements

This work was funded by the Väisälä project grant RESOLVE by the Finnish Academy of Science Letters (M.B.), the Jane and Aatos Erkko Foundation grant ROOTS (M.B.), as well as Svenska Kulturfonden (M.B.). The authors acknowledge the provision of facilities and technical support by Aalto University at OtaNano-Nanomicroscopy Center (Aalto-NMC).